\documentclass[twocolappendix]{emulateapj}
\usepackage{amsmath}
\usepackage[]{graphicx}
\usepackage{bm}
\usepackage{color}
\usepackage{mathrsfs}
\usepackage{appendix}

\begin{document}

\title{Exclusion of cosmic rays from molecular clouds by self-generated electric fields }

\author{Kedron Silsbee$^1$, Alexei V. Ivlev$^1$}
\email[e-mail:~]{ksilsbee@mpe.mpg.de} \email[e-mail:~]{ivlev@mpe.mpg.de} \affiliation{$^1$Max-Planck-Institut f\"ur
Extraterrestrische Physik, 85748 Garching, Germany }

\begin{abstract}
It was recently discovered that in some regions of the Galaxy, the cosmic ray (CR) abundance is several orders of magnitude
higher than previously thought. Additionally, there is evidence that in molecular cloud envelopes, the CR ionization may be
dominated by electrons. We show that for regions with high, electron-dominated ionization, the penetration of CR electrons
into molecular clouds is modulated by the electric field that develops as a result of the charge they deposit. We evaluate
the significance of this novel mechanism of self-modulation and show that the CR penetration can be reduced by a factor of a
few to a few hundred in high-ionization environments, such as those found near the Galactic center.
\end{abstract}


\maketitle

\section{Introduction}

Understanding the transport of cosmic rays (CRs) in dense gas is one of the big open questions of astrophysics. Low-energy
CRs govern the evolution of molecular clouds and the formation of stars \citep{Caselli12, Padovani20}, being the dominant
source of ionization \citep{McKee89, Keto08, Neufeld17} and UV emission \citep{Prasad83} above a column density of
$\sim 10^{22}$~cm$^{-2}$. These processes affect both the chemistry \citep{Keto08, Keto14} and thermodynamics
\citep{Galli2002, Glassgold2012, Ivlev19} of the clouds. Furthermore, the level of ionization governs the degree to which
the gas is coupled to the magnetic field \citep{Shu1987}. This has profound implications for the existence and size of disks
around young stars \citep{Zhao16, Zhao18}.

For a long time the CR ionization rate $\zeta$ was thought to likely be on the order of $10^{-17}$~s$^{-1}$, based on
measurements of the CR abundance near Earth \citep{Spitzer68}. More recently there have been measurements of $\zeta$ in
nearby molecular clouds \citep{Indriolo12}, suggesting $\zeta$ as high as $10^{-15}$~s$^{-1}$ towards some
clouds.  In some environments the CR ionization rates can be orders of magnitude higher still. In the Central Molecular
Zone (CMZ) of the Galaxy, \citet{LePetit16} and  \citet{Oka19} estimate $\zeta$ of $1-11\times 10^{-14}$~s$^{-1}$ and $2
\times 10^{-14}$~s$^{-1}$ respectively.  \citet{Yusefzadeh07} suggest a rate of $5 \times 10^{-13}$~s$^{-1}$
in the Sagittarius C region. There is also evidence of extremely high CR abundance near young stars \citep{Ceccarelli14,
Ainsworth14}.

It is not known whether CR protons or electrons are the primary source of ionization.  As shown in \citet{Padovani18},
if the spectra of electrons and protons measured by the Voyager probes \citep{Stone19} are extrapolated down to lower
energies, then $\zeta$ is dominated by electrons at column densities lower than $2 \times 10^{21}$~cm$^{-2}$.  If the
CR electron and proton spectra have the slope appropriate for acceleration in strong shocks, then ionization is dominated by
electrons unless protons dominate the total CR energy by factors of tens.

The commonly used free-streaming model for the CR transport in clouds and disks \citep{Padovani09, Padovani18} holds that
they propagate along local magnetic field lines without substantial pitch-angle scattering, and lose energy due to
interactions with the gas in the cloud (losses are dominated by ionization for non-relativistic particles). The fundamental
effect completely neglected in such models (also those including CR scattering on magnetic disturbances) is
an inevitable net deposition of charge within the cloud. Low-energy CRs are absorbed in the
cloud, becoming thermalized charged particles. So as to maintain charge balance within the cloud, the thermal plasma must
transport a net current.  Because the plasma has a finite conductivity, this implies the presence of a long-range electric
field which acts to modulate the penetration of CRs.

This mechanism of CR modulation -- which has not been considered thus far, to our knowledge -- is the topic of the present
Letter. We show that the self-generated electric field is strong enough to have a large effect for reasonable parameters,
provided CR electrons dominate the ionization.

\section{Linear Regime}

Let us calculate the steady-state electric potential in a cloud in the limit that the incoming CR flux is not modulated by
the electric field. We approximate the cloud as a slab of a weakly ionized cool gas with uniform density $n$, embedded in a
warm infinitely conducting medium filled with CRs. The magnetic field lines are assumed to be straight, but may enter the
cloud at an arbitrary angle with respect to the surface. The column density $N$ relevant to the CR propagation is defined by
integrating the density {\it along the magnetic field}. The distance $z$ is measured in this direction, too, and set to 0 at
the cloud center, where $N$ is half of the total cloud value $N_{\rm cl}$. For typical diffuse clouds (even with the extreme
ionization implied by our model), the plasma conductivity parallel to the magnetic field is higher by at least 6 orders of
magnitude than the perpendicular conductivity. Hence, the current due to charge deposition by CRs in the region between $0$
and $z$ within the cloud must be simply balanced by the parallel plasma current at position $z$.

Keeping in mind that the cloud is bombarded by CRs from both sides, the CR current $J_{\rm CR}$ at column
$N$ is given by the integral
\begin{equation}
J_{\rm CR}(N)= 2 \pi q_{\rm CR} \int_0^1d\mu\:\mu \int_{\mathscr{E}_{\rm ext}(N/\mu)}^{\mathscr{E}_{\rm ext}
\left([N_{\rm cl} - N]/\mu\right)}d\mathscr{E}\: j_i(\mathscr{E}),
\label{eq:deposition}
\end{equation}
where $q_{\rm CR}=\pm e$ is the charge of a CR particle and $\mu$ is the cosine of the pitch angle. The integration limits
are determined by the extinction energy $\mathscr{E}_{\rm ext}(N)$ -- the lowest energy of a particle that can penetrate to
column depth $N$. The external (initial) spectrum of CRs, $j_i(\mathscr{E})$, is assumed to be isotropic and given by
\begin{equation}
j_i(\mathscr{E}) = j_0 \left(\frac{\mathscr{E}_0}{\mathscr{E}}\right)^{a}~{\rm s}^{-1}\,{\rm cm}^{-2}\,
{\rm eV}^{-1}\,{\rm sr}^{-1}.
\label{eq:spectrum}
\end{equation}
To determine $\mathscr{E}_{\rm ext}(N)$, we must introduce the ionization loss function. This is given in \citet{Padovani18}
as
\begin{equation}
L \equiv \frac{d\mathscr{E}}{dN} = L_0 \left(\frac{\mathscr{E}_0}{\mathscr{E}}\right)^{s},
\label{eq:lossFunction}
\end{equation}
and then
\begin{equation}
\mathscr{E}_{\rm ext}(N) = \mathscr{E}_0\left(\frac{N}{N_0}\right)^\frac{1}{1+s},
\label{eq:Eext}
\end{equation}
where $N_0 = (1+s)^{-1}\mathscr{E}_0/L_0$.
Performing the integral in Equation \eqref{eq:deposition} yields
\begin{equation}
J_{\rm CR}(N) = \frac{2\pi(1+s) q_{\rm CR}j_0\mathscr{E}_0}{(1-a)(1+2s+a)} \left[(\tilde N_{\rm cl}-\tilde N)^{\frac{1-a}{1+s}}
- \tilde N^{\frac{1-a}{1+s}}\right],
\label{eq:specialDeposition0}
\end{equation}
where $\tilde N \equiv N/N_0$. For $a = 1$:
\begin{equation}
J_{\rm CR}(N) = \frac{\pi q_{\rm CR} j_0 \mathscr{E}_0}{1+s} \ln{\left(\frac{N_{\rm cl} - N}{N}\right)}.
\label{eq:specialDeposition}
\end{equation}
If we are interested in column densities between $10^{19}$ and $10^{22}$~cm$^{-2}$, the relevant energies are from 5 to
300~KeV for electrons, and from 100~KeV to 8~MeV for protons. The loss functions on these intervals are well approximated by
Equation \eqref{eq:lossFunction} with $s = 0.75$, $L_0 = 2.0 \times 10^{-16}$~eV~cm$^{2}$, $\mathscr{E}_0 =10$~KeV for electrons,
and $s = 0.78$, $L_0 = 3.7 \times 10^{-16}$~eV~cm$^{2}$, $\mathscr{E}_0=10$~MeV for protons. Even though $s$ may change a
little, depending on the energy range, these variations have negligible impact on our results. Therefore, in this Letter we
employ the above values for numerical calculations, while keeping $s$ explicitly in the analytical results. As in
\citet{Padovani18}, we use the number density of all gas particles, rather than of hydrogen atoms, and assume the hydrogen
to be molecular.

Let us denote with $E$ the electric field component {\it parallel} to the magnetic field. The value of
$E(N)$ in the cloud is obtained from the the steady-state condition
\begin{equation}\label{E_field}
J_{\rm CR}(N)+J_{\rm pl}(E)=0,
\end{equation}
where $J_{\rm pl}= \sigma E$ is the plasma current along the magnetic field, determined by the corresponding electric
conductivity \citep{Braginskii65},
\begin{equation}
\sigma = 5\times 10^{9}~{\rm s}^{-1} \left(\frac{T}{50~{\rm K}}\right)^{3/2}.
\label{eq:sigma}
\end{equation}
Equation \eqref{eq:sigma} assumes a fully ionized plasma.  The expected ionization fraction in the outer layers of a
cloud is in excess of $10^{-3}$ \citep{Neufeld17}.  Since the electron-neutral collision cross section is lower by 5 or
6 orders of magnitude than the electron-ion cross section for such conditions, the neutrals have a negligible effect on the
parallel conductivity.

Consider the CR spectra measured from the Voyager probes \citep{Stone19} and extrapolated to lower energies as in
\citet{Padovani18}, and a cloud with $N_{\rm cl} = 6\times 10^{21}$~cm$^{-2}$, $n =60$~cm$^{-3}$, and $T = 50$~K
\citep{Draine11}. From Equations \eqref{eq:specialDeposition0}, \eqref{E_field}, and \eqref{eq:sigma} we derive
the magnitude of the electric potential energy $e|\phi(N)|$ for CR protons ($p$) and electrons ($e$), and compare these
with the respective extinction energies $\mathscr{E}_{\rm ext}(N)$. We obtain that $e\phi_p$ is completely negligible at
any $N$, while $e|\phi_e(N)|$ is just a factor of 30 less than $\mathscr{E}_{{\rm ext},e}(N)$. Since the electron spectrum
dominates the ionization at low column density, this implies that if the CR abundance were increased by a factor of 30, then
$\zeta$ would be significantly affected by the electron charge buildup. Such an increased spectrum would result in a total
ionization rate of somewhat less than $10^{-15}$~s$^{-1}$ at a column density of $10^{21}$~cm$^{-2}$. This value is within
the range estimates made by \citet{Indriolo12}, suggesting that this effect may play a role, even in local molecular clouds.

The importance of the electric field (at a given $\zeta$) is substantially higher if the ionization is dominated
by electrons.  Consider locations adjacent to a strong shock which is acting as a source of CRs, so one can expect
$\mathscr{F}(p)\propto p^{-2}$ for the particle density in {\it momentum space}. Assuming an electron to proton ratio of
$\chi$, and column densities such that the ionization is dominated by non-relativistic particles (up to a few times
$10^{22}$~cm$^{-2}$ for electrons, and a few times $10^{25}$~cm$^{-2}$ for protons), we obtain the {\it energy} spectra
given by Equation \eqref{eq:spectrum} with $j_{0,e}/j_{0,p} =\chi m_p/m_e$. Then Equation \eqref{eq:specialDeposition} shows that the
deposited charge is dominated by electrons if $\chi > m_e/m_p$. From Equation 31 of \citet{Silsbee19}, we find that the ratio
of the ionization rates at a given column density is $\zeta_e/\zeta_p\sim \chi(m_p/m_e)^{s/(1+s)}$ (for simplicity, we
assume the same $s$ for electrons and ions and set $L_{0,e}/L_{0,p}\sim m_e/m_p$). From this we conclude that, if $\chi$ is
greater than a few percent, then both the ionization and the charge deposition are dominated by electrons.

Studies of particle acceleration are still uncertain as to the value of $\chi$ -- there is evidence that it is less than 1\%
in quasi-parallel shocks \citep{Park15}. On the other hand, there is recent evidence \citep{Spitkovsky19} that
quasi-perpendicular shocks in fact preferentially accelerate electrons. Hence, it is reasonable to assume
that there are regions where more than a few \% of the electrons have been produced in quasi-perpendicular shocks, and
therefore for the remainder of this Letter, we consider regions in which CR electrons dominate the ionization rate.

\section{High-flux limit for CR electrons}

We now approach the problem from a different perspective. Instead of assuming the electric field to be a small perturbation
on the propagation of CRs, we consider it to be the dominant effect and treat ionization losses as a perturbation. To be
more precise, we assume that at every position of interest within the cloud, the electric potential $\phi$ satisfies
$e|\phi|\gg \mathscr{E}_{\rm ext}$.

Let us consider, as before, a slab of uniform gas with a constant angle between the magnetic field and the
surface. Now the distance coordinate $z$, measured along the magnetic field, is set to be 0 at one edge of the cloud. We
posit that the absolute value of the potential as a function of $z$ over some range of distances is given
by
\begin{equation}
|\phi(z)| = \frac{\mathscr{E}_0}e \left(\frac{z}{z_0}\right)^f,
\label{eq:phi}
\end{equation}
with the length scale $z_0$ and exponent $0<f<1$ to be determined. Obviously since the electric field changes direction at
the center of the cloud, this form for $\phi(z)$ is not valid near the center of the cloud, so we restrict our attention to
$z$ much smaller than the cloud size. This allows us to solve for $E$, the electric field component parallel
to the magnetic field, as a function of position:
\begin{equation}
E(z) = E_0\left(\frac{z}{z_0}\right)^{f-1},
\label{eq:Efield}
\end{equation}
where $E_0=f\mathscr{E}_0/(ez_0)$.

Instead of using variables $\mathscr{E}$ and $\mu$, for our problem below we find it more convenient to work with the
``parallel'' and ``perpendicular'' energies, $\mathscr{E}_\|= \mathscr{E} \mu^2$ and $\mathscr{E}_\perp= \mathscr{E}
(1-\mu^2)$. The local spectrum per unit $\mathscr{E}_\|$ and $\mathscr{E}_\perp$ can be conveniently calculated from the
local density in the momentum space. According to the Liouville theorem, the CR density in momentum space is
conserved along the phase trajectories, that is to say $\mathscr{F}({\bf p},{\bf r})=\mathscr{F}_i(p^2+ 2me|\phi({\bf
r})|)$. Combining this with a general relation $j(\mathscr{E}, \mu) = p^2\mathscr{F}(p, \mu)$, we find
$\mathscr{F}(p,z)=j_i(\mathscr{E}+e|\phi|)/[2m(\mathscr{E}+e|\phi|)]$. Then, noting that $2\pi p_\perp dp_\perp dp_\|= \pi
m\sqrt{2m/\mathscr{E}_\|}\: d\mathscr{E}_\|d\mathscr{E}_\perp$, we multiply $\mathscr{F}(p,z)$ with the physical velocity
$\sqrt{2\mathscr{E}/m}$ and the pre-factor $\pi m\sqrt{2m/\mathscr{E}_\|}$, which yields the spectrum expressed in new
variables,
\begin{equation}\label{local_2}
j(\mathscr{E}_\|, \mathscr{E}_\perp, z)=\pi \sqrt{\frac{\mathscr{E}}{\mathscr{E}_\|}}\:
\frac{j_i(\mathscr{E}+e|\phi|)}{(\mathscr{E}+e|\phi|)}\:,
\end{equation}
where we use $\mathscr{E}=\mathscr{E}_\|+\mathscr{E}_\perp$ for brevity.

As the first step, we equate the current of CRs which is absorbed beyond position $z$ due to the losses and the
plasma current along the magnetic field. Integrating over the {\it initial} CR distribution at the cloud
edge, we obtain
\begin{equation}
\frac{\sigma E(z)}{e} = \int_{e|\phi(z)|}^\infty d\mathscr{E}_\parallel\int_0^{\mathscr{E}_\perp^{\rm cr}(\mathscr{E}_\|)}
d\mathscr{E_\perp}\:j(\mathscr{\mathscr{E}_\parallel, \mathscr{E}_\perp},0),
\label{eq:master}
\end{equation}
with $\sigma$ from Equation \eqref{eq:sigma}. Particles with $\mathscr{E_\perp} = 0$ will have zero kinetic energy at the
turning point, and will therefore be stopped and contribute to the charge buildup. Particles with relatively large
$\mathscr{E_\perp}$ have enough transverse energy that they are accelerated back to the cloud edge before their energy is
damped. Hence, for particles with a given $\mathscr{E}_\parallel$ there is a critical value of $\mathscr{E}_\perp$, denoted
$\mathscr{E}_\perp^{\rm cr}$, which determines their trapping inside the cloud. As discussed in Appendix A, the dynamics of
a particle with initial transverse energy $\mathscr{E}_{\perp}$ in the presence of losses are determined by a dimensionless
number
\begin{equation}
M = \frac{eE_{\rm turn}}{n L(\mathscr{E}_{\perp})}\:,
\label{eq:M}
\end{equation}
where $E_{\rm turn}$ is the parallel electric field at the turning point.  The critical initial transverse
energy $\mathscr{E}_\perp^{\rm cr}$ for $\mathscr{E}_\|=e|\phi(z)|$ corresponds to $M_{\rm cr}\approx3.6$: For $M<M_{\rm
cr}$, particles are stopped by the losses near the turning point, otherwise they return back to the cloud edge. Using
Equations \eqref{eq:phi} and \eqref{eq:Efield}, we find that the electric field at the turning point is
\begin{equation}
E_{\rm turn} = E_0\left(\frac{\mathscr{E}_0}{\mathscr{E}_\parallel}\right)^\frac{1-f}{f}.
\label{eq:Eturn}
\end{equation}
Combining Equations \eqref{eq:M} and \eqref{eq:Eturn}, we find
\begin{equation}
\mathscr{E}_\perp^{\rm cr} = \mathscr{E}_0 \left(M_{\rm cr}\frac{nL_0}{eE_0}\right)^{\frac1s}
\left(\frac{\mathscr{E}_\parallel}{\mathscr{E}_0}\right)^{\frac{1-f}{sf}}.
\label{eq:EPerpCrit}
\end{equation}
Plugging Equation \eqref{eq:EPerpCrit} into Equation \eqref{eq:master} and approximating that $\mathscr{E}_\perp^{\rm cr}
\ll \mathscr{E}_\parallel$ (which is verified in Section~\ref{notes}), we can evaluate the integral in Equation
\eqref{eq:master} under condition $f(1+as)>1$. This yields
\begin{align}
& \frac{\sigma E_0}{e} \left(\frac{z_0}{z}\right)^{1-f} =  \pi j_0\mathscr{E}_0 \left(M_{\rm cr}\frac{nL_0}{eE_0}\right)^{\frac1s}
\nonumber \\
& \times\frac{sf}{f(1+as)-1} \left(\frac{z_0}{z}\right)^ \frac{f(1+as)-1}{s}.
\end{align}
Matching powers of $z$, we find
\begin{equation}
f = \frac{1+s}{1+s+as}\:.
\label{eq:csol}
\end{equation}
This allows us to solve for $z_0$:
\begin{equation}
\frac{nz_0}{N_0} = \left(\frac{(1+s)^{\frac1s-1}f^{\frac1s+1}}{M_{\rm
cr}^{\frac1s}}\:
\frac{as\sigma n}{\pi e^2j_0N_0}\right)^{\frac{s}{1+s}}. \label{eq:phi0}
\end{equation}
We note that the condition $f(1+as)>1$ is reduced to $a>0$. We finally derive
\begin{equation}\label{phi_to_E}
\frac{e|\phi|}{\mathscr{E}_{\rm ext}}=\left(\frac{N_0}{nz_0}\right)^{\frac{1+s}{1+s+as}}
\left(\frac{N}{N_0}\right)^{\frac{s(1+s-a)}{(1+s)(1+s+as)}},
\end{equation}
naturally, invariant with respect to the choice of $\mathscr{E}_0$.
We require $e |\phi|/\mathscr{E}_{\rm ext} \gg 1$ in order for the solution in Equation \eqref{phi_to_E} to be valid. Formally, it
must break down either at high or low $N$, depending on the sign of the slope. In fact, however, the slope is very small:
$\approx0.13$ ($-0.03$) for $a = 1$ ($a = 2$). For this reason, as a practical matter, over the range of column density of
interest the solution either applies everywhere, or applies nowhere -- depending on the magnitude of $nz_0/N_0$, which is
the chief parameter characterizing the effect of self-generated field.

Now we can calculate the ionization rate in the high-flux limit. Again, we assume that the regular
(ionization) losses play no role in determining the local CR spectrum, which is determined purely by the external spectrum
and the electric potential. The primary CR ionization rate of H$_2$ at position $z$ is given by
\begin{equation}
\zeta_\phi(z) = \frac{2}{\epsilon} \int_0^\infty \int_0^\infty d\mathscr{E}_\|d\mathscr{E}_\perp\:
j(\mathscr{E}_\parallel, \mathscr{E}_\perp, z) L(\mathscr{E}),
\end{equation}
where $j(\mathscr{E}_\parallel, \mathscr{E}_\perp, z)$ is given by Equation \eqref{local_2} and $\epsilon$ is the mean energy
lost per primary ionization event \citep{Silsbee19}, which we take to be 58 eV. We obtain
\begin{equation}
\zeta_\phi = \frac{4 \pi Bj_0 L_0\mathscr{E}_0}{\epsilon} \left(\frac{e|\phi|}{\mathscr{E}_0}\right)^{-(a+s-1)},
\label{eq:finalZeta}
\end{equation}
where $B\equiv B(2-s,a+s-1)$ is the beta function (see Appendix B). By comparing Equation \eqref{eq:finalZeta} with
Equation (31) of \citet{Silsbee19}, which describes the ``regular'' CR ionization rate $\zeta(N)$, we obtain
\begin{equation}\label{zeta_ratioText}
\frac{\zeta_\phi}{\zeta}\approx1.7\left(\frac{e|\phi|}{\mathscr{E}_{\rm ext}}\right)^{-(a+s-1)},
\end{equation}
where $e|\phi|/\mathscr{E}_{\rm ext}$ is given by Equation \eqref{phi_to_E} and the pre-factor is accurate within 3\% for $1\leq
a\leq2$, see Equation (B3).

We point out that the sign of $a+s-1$ in Equations \eqref{eq:finalZeta} and \eqref{zeta_ratioText} coincides with the sign of the
exponent which determines the regular dependence $\zeta(N)$, Equation 33 of \citet{Silsbee19}. In case $a+s-1<0$ the CR
spectrum is too hard and low-energy particles are no longer dominating ionization. Hence, as for the case of regular
ionization, $a+s-1>0$ is assumed.

\subsection{Magnitude of the effect}

Using Equations \eqref{eq:lossFunction}, \eqref{eq:sigma}, and \eqref{eq:finalZeta}, we rewrite Equation \eqref{phi_to_E} in
terms of the physical parameters:
\begin{equation}\label{practical}
\frac{e |\phi|}{\mathscr{E}_{\rm ext}} \approx 1.3\: \exp{(0.35a)}\:T_{250}^{-0.58}\:N_{21}^{0.33}\:(\zeta_{-15}/n_{30})^{0.39},
\end{equation}
with $T_{250}$ in units of 250~K, $N_{21}$ in units of $10^{21}$~cm$^{-2}$, $\zeta_{-15}$ (evaluated at same column
density) in units of $10^{-15}$~s$^{-1}$, and $n_{30}$ in units of $30~{\rm cm}^{-3}$. Equation \eqref{practical} is
accurate to within 2.5\% for $1\leq a\leq2$.

As an example, we consider conditions appropriate for the high ionization rate regions near the Galactic center.  We assume
that these ionization rates are dominated by CR electrons, and consider an electron spectrum with $a = 1$ (appropriate for
acceleration by strong shocks) and $a = 2$ (for weaker shocks with compression ratio of 2) \citep{Blandford78}. In both
cases, we choose $j_0$ so that (with the electric field included) the primary ionization rate at
$N=10^{21}$~cm$^{-2}$ is equal to $4 \times 10^{-14}$~s$^{-1}$, based on the values of $1-11 \times 10^{-14}$~s$^{-1}$
reported in \citet{LePetit16} for the CMZ.
\begin{figure}[htp]
\centering
\includegraphics[width=1.03\columnwidth]{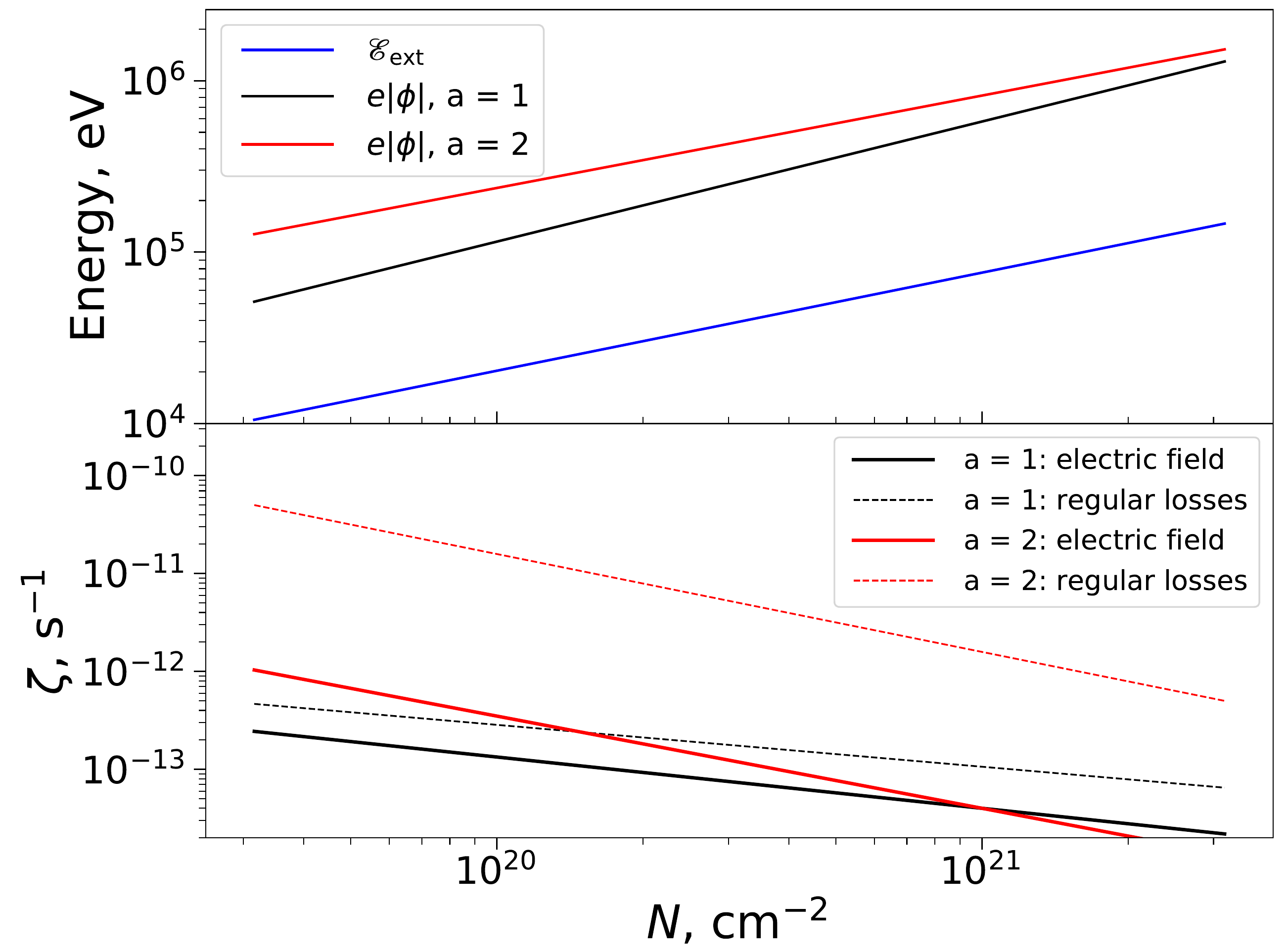}
\caption{The top panel shows a comparison between the electric energy $e|\phi|$ [Equation \eqref{phi_to_E}] and the extinction
energy $\mathscr{E}_{\rm ext}$ [Equation \eqref{eq:Eext}] for CR electrons. The black and red curves are for an electron spectrum
with $a = 1$ and 2, respectively (normalized such that $\zeta = 4 \times 10^{-14}$ at $N = 10^{21}$~cm$^{-2}$). The bottom
panel shows the corresponding ionization rate, plotted without and with taking into account the self-generated electric
field [Equation 31 of \citet{Silsbee19} and Equation \eqref{eq:finalZeta} of this Letter, respectively].}
\label{ionizationRates}
\end{figure}

The top panel of Figure~\ref{ionizationRates} shows a comparison of $e|\phi(N)|$ with $\mathscr{E}_{\rm ext}(N)$. We use $T
= 250$~K and $n = 30$~cm$^{-3}$, based on the observations in \citet{LePetit16}. The bottom panel shows a comparison of the
ionization rate calculated from Equation \eqref{eq:finalZeta} with that calculated ignoring electric fields (Equation 31 of
\citet{Silsbee19}).  At a representative column density of $10^{21}$~cm$^{-2}$, $e|\phi|/\mathscr{E}_{\rm ext}\approx8$ for
the spectrum with $a = 1$, and $\approx11$ for $a = 2$, leading to reductions in the ionization rate by factors of about 2.7
and 40, respectively. Note that for $T\approx50$~K, suggested by Figure 9 of \citet{Bisbas15} for our values of $\zeta$, the
reduction would be about 5.3 and 200, respectively.

Limits of the non-relativistic formulation are reached at higher column densities, where $e|\phi| \gtrsim mc^2$. In
Appendix C, the calculations presented in Equations \eqref{local_2}--\eqref{phi_to_E} are redone in the
ultra-relativistic regime, assuming that outside the cloud the CR density in momentum space has the same power-law slope for
both relativistic and non-relativistic particles. It is also assumed that the critical kinetic energy is still
non-relativistic near the turning point -- this premise is shown to be valid for $N$ substantially higher than
$10^{22}$~cm$^{-2}$, i.e., well applicable for molecular clouds. We find that the electric potential \eqref{phi_to_E} is
modified in the ultra-relativistic regime as
\begin{equation}
\phi_{\rm rel}(N) = \phi(N)\left(\frac{N}{N_{\rm rel}}\right)^{-\frac{s(1+a)(1+s)}{(1+s+as)(1+2s+2as)}},
\end{equation}
Here $N_{\rm rel}$ is the column density at which $e|\phi(N)|$ from Equation \eqref{phi_to_E} is equal to $\approx2.4mc^2$, see
Equation (C6). For Figure \ref{ionizationRates}, $N_{\rm rel} =2-3\times 10^{21}~{\rm cm^{-2}}$ and the exponent
varies between $-0.26$ and $-0.22$ for $1\leq a\leq2$. Hence, relativistic effects only lead to a minor modification of the
self-generated electric field, and the CR modulation remains essentially unchanged.

\subsection{Notes on the Derived Solution}
\label{notes}

Here we verify important assumptions made to derive the above results, and briefly discuss some immediate implications.

\subsubsection{Anisotropy}

We can check the assumption made after Equation \eqref{eq:EPerpCrit}, that $\mathscr{E}_\perp^{\rm cr} \ll
\mathscr{E}_\parallel$. Setting $e|\phi|\approx \mathscr{E}_\|$, we rewrite Equation \eqref{eq:EPerpCrit} as
\begin{equation}
\frac{\mathscr{E}_\perp^{\rm cr}}{\mathscr{E}_\|} \approx \left(\frac{M_{\rm cr}}{f(1+s)}\right)^\frac{1}{s}
\left(\frac{\mathscr{E}_{\rm ext}}{e|\phi|}\right)^{\frac1s+1}.
\label{eq:anisotropy}
\end{equation}
For the parameters in Figure \ref{ionizationRates}, the anisotropy due to CR deposition is expected to be less than a
few \%. This means that the excitation of the streaming instability by penetrating CRs will be suppressed compared to a case
of no electric field \citep{Morlino15, Ivlev18} (where the pitch angle anisotropy could be of order unity).

\subsubsection{Electric Fields from Alfv{\'e}n Waves}

We have implicitly assumed that the deposition of CRs is the only source of a large-scale electric field parallel to
the magnetic field. In fact, an electric field could also be produced by Alfv{\'e}n waves present in the cloud as a result
of turbulence. In ideal MHD, such fields are perpendicular to the local magnetic field, and thus do no work on CRs. It is
worth noting though that the strength of this electric field, associated with turbulent motions with the velocity $u$, is on
the order of $uB/c\sim 10^{-11}$~statV~cm$^{-1}$ for typical parameters. This is $\sim10^5$ times stronger than the field
from the modulation effect, and therefore it could be significant if there were even a very small deviation from
orthogonality.

In particular, \citet{Bian10} and \citet{Klimushkin14} suggest that under realistic conditions, Alfv{\'e}n waves are
able to generate a small parallel electric field, but its strength is highly uncertain. The authors developed a model under
which the ratio of parallel to perpendicular electric field is roughly the squared ratio of the ion gyroradius to the
wavelength. Estimates of the lower wavelength bound for Alfv{\'e}n waves, given in Appendix~C of \citet{Kulsrud69}, suggest
that waves are cut off around the ambipolar damping scale of $\sim10^{16}$~cm for our conditions. For typical magnetic
fields, the ion gyroradius is a few times $10^6$~cm, so the parallel electric field arising from such mechanism is smaller
than the perpendicular field by some 19 orders of magnitude.

We are not aware of a model which shows a significant parallel electric field in the long-wavelength limit. However, as
both the cutoff scale of Alfv{\'e}n waves and the magnitude of parallel electric field are subject to significant
uncertainties, this could be an interesting avenue of future work.

\subsubsection{Joule Heating}

The electric current $J_{\rm pl}=\sigma E$ induced in the gas due to CR deposition represents an additional source of
gas heating.
The rate of the resulting Joule heating is
\begin{equation}
\label{eq:HJ}
 H_J = \sigma E^2.
\end{equation}
This should be compared to the rate of regular gas heating by CRs, given by
\begin{equation}
H_{\rm CR} = \eta \epsilon \zeta_{\phi}n,
\label{eq:HI}
\end{equation}
where $\eta$ is an efficiency factor of order 40\% \citep{Glassgold12}. As shown in Appendix D, their ratio is
\begin{equation}
\frac{H_J}{H_{\rm CR}} = Q \left(\frac{e|\phi|}{\mathscr{E}_{\rm ext}}\right)^{-\frac1{s}+s},
\end{equation}
where $Q$ varies between $\approx7$ and $\approx10$ for $1\leq a \leq2$. Thus, for conditions illustrated in Figure~\ref{ionizationRates}, Joule heating is larger than the regular CR heating by a factor which varies monotonically between 2.0 and 2.4 for $1 \leq a \leq 2$.  Regardless of $a$, Joule heating becomes subdominant in the limit of very strong self-modulation.

\section{Conclusion}

We propose a novel mechanism of CR self-modulation, which can substantially reduce the penetration of CR electrons into
molecular clouds. The penetration is limited by the electric fields generated due to the deposition of those same electrons.
If the electron spectrum is produced by acceleration in strong shocks, the ionization rate can be reduced by a factor of a
few in the high-ionization environments found in our Galaxy, such as CMZ. The reduction becomes much stronger for steeper
spectra, appropriate for weaker shocks. Hence, the high ionization rates near the Galactic center could imply even higher CR
energy densities than previously thought. The effect is more pronounced at lower gas number densities, where direct
measurements of the ionization can be made. The ionization rate in denser regions will therefore be much higher than would
be predicted from measurements coupled with conventional models of CR transport.  Furthermore, the electric current
induced in the gas due to the CR deposition represents an additional heating source. We show that the resulting Joule
heating could be of similar magnitude to the regular gas heating by CRs.

\bibliographystyle{apj}
\bibliography{letter}

\appendix
\section{Appendix A: Calculation of $\mathscr{E}_\perp^{\rm cr}(\mathscr{E}_\parallel)$}

As stated in the main text, we consider the limit $e |\phi|\gg \mathscr{E}_{\rm ext}$. In this case only particles with
sufficiently small $\mathscr{E}_{\perp}$ can be trapped in a cloud, as they are slowed nearly to a stop at the turning
point, thus suffering strong ionization losses. To estimate $\mathscr{E}_\perp^{\rm cr}(\mathscr{E})$, we use the loss
function given by Equation (3). We are dealing with non-relativistic particles, such that their velocities are $v
=\sqrt{2\mathscr{E}/m}$, where $m$ is the particle mass.

We note that for a particle moving parallel to an electric field with strength $E$, there is a critical energy
$\mathscr{E}_{\rm cr}$ such that the drag force $nL(\mathscr{E}_{\rm cr})$ due to energy losses is compensated by the
acceleration from the electric field,
\begin{equation}
\mathscr{E}_{\rm cr} = \mathscr{E}_0 \left(\frac{nL_0}{eE}\right)^{\frac1s}.
\end{equation}
We assume that if $\mathscr{E}_{\rm cr}$ is reached after turnaround, this occurs in a short distance from the turning
point, and we can therefore calculate $\mathscr{E}_\perp^{\rm cr}$ assuming a constant electric field. This assumption is
verified at the end of our calculation.

In a constant electric field $E$, the equations of motion for the parallel and transverse velocities are:
\begin{equation}
m\dot v_\parallel=-eE -nL(\mathscr{E})\frac{v_\parallel}{v}\;, \quad\quad
m\dot v_\perp=-nL(\mathscr{E})\frac{v_\perp}{v}\;.
\end{equation}
Normalizing the velocity to the initial transverse velocity, $\tilde v=v/v_{\perp i}$, and time to
\begin{equation}
\tau = \frac{mv_{\perp i}}{n L(\mathscr{E}_{\perp i})}\;,
\end{equation}
we then arrive at the equations
\begin{equation}
\dot{\tilde v}_\| = -M - \tilde v_\| \tilde v^{-2s-1}, \quad\quad
\dot{\tilde v}_\perp = -\tilde v_\perp \tilde v^{-2s-1},
\label{eq:coupledMotion}
\end{equation}
containing a single dimensionless number
\begin{equation}
M = \frac{eE}{n L(\mathscr{E}_{\perp i})}\;.
\end{equation}
To distinguish between the local and initial values, here we identify the latter with the subscript $i$. The numerical
solution of Equations \eqref{eq:coupledMotion} shows that for $M < M_{\rm cr} \approx3.6$ particle trajectories decay to zero
velocity. Furthermore, we find that for $M/M_{\rm cr} \leq 0.999$, the final position of the particle relative to the
turning point $z_{\rm turn}$ satisfies $\Delta z < 0.2\:v_{\perp i}\tau = 0.4M\:\mathscr{E}_{\perp i}/(eE)$.  Noting that
the turning point occurs approximately where $e|\phi(z)|=\mathscr{E}_{\|i}$ and using Equations (9) and (10),
we find $z_{\rm turn} = f\mathscr{E}_{\|i}/ (eE)$, and
\begin{equation}
\frac{\Delta z}{z_{\rm turn}} < \frac{0.4M}{f} \frac{\mathscr{E}_{\perp i}}{\mathscr{E}_{\|i}}\;.
\end{equation}
As shown in Equation (25), in the limit $e|\phi|\gg \mathscr{E}_{\rm ext}$ we have $\mathscr{E}_\perp^{\rm cr}/ \mathscr{E}_\| \ll
1$ and, hence, $\Delta z/z_{\rm turn} \ll 1$. Thus, the assumption that trapped particles are stopped near the turning point
is justified.

\section{Appendix B: Derivation of Equations (21) and (22)}

We substitute $L(\mathscr{E})$ and $j(\mathscr{E}_\parallel, \mathscr{E}_\perp, z)$, given by Equations (3) and (11), to Equation (20)
and, normalizing energies by $e|\phi|$, readily obtain Equation (21) with the pre-factor proportional to the following double
integral:
\begin{equation}
I(a,s)=\int_0^\infty \int_0^\infty dpdq\:\frac{(p+q)^{1/2-s}}{\sqrt{p}(p+q+1)^{a+1}}\:.
\end{equation}
We replace the integration variables $p,q$ by $x^2,y^2$ and rewrite the integral in polar coordinates with
$r=\sqrt{x^2+y^2}$ and $\tan\theta=y/x$. Integrating over $\theta$ between 0 and $\pi/2$ and then substituting
$r=\sqrt{t/(1-t)}$ yields
\begin{equation}
I(a,s)=2B(2-s,a+s-1),
\end{equation}
expressed via the beta function.

Equation (22) is derived by comparing Equation (21) with Equation 31 of \citet{Silsbee19}, which describes the ``regular''
CR ionization rate $\zeta(N)$. We note that Equation 31 should be multiplied by $2\pi$, due to a different normalization of
the CR spectrum in \citet{Silsbee19}, and $d$ denotes $s$ in our Letter.  Also, unlike \citet{Silsbee19}, we assume that the
magnetic field has constant strength, so there is no magnetic mirroring \citep{Silsbee18}.  The integral $I_f$, entering
Equation 31 and given by Equation 32 of \citet{Silsbee19}, can also be expressed via the beta function, by substituting
$x^{1+s}=t/(1-t)$ for the integration variable.
Using Equation (4), we finally obtain
\begin{equation}\label{zeta_ratio}
\frac{\zeta_\phi}{\zeta}=2(a+2s)\:\frac{B(2-s,a+s-1)}{B\left(\frac1{1+s},\frac{a+s-1}{1+s}\right)}
\left(\frac{e|\phi|}{\mathscr{E}_{\rm ext}}\right)^{-(a+s-1)}.
\end{equation}
The pre-factor of $(e|\phi|/\mathscr{E}_{\rm ext})$ is a slowly varying function of $a$, equal to $\approx1.7$ for
$1\leq a\leq2$.

\section{Appendix C: The ultra-relativistic regime}

We calculate the electric potential in the limit $e|\phi| \gg mc^2$, assuming that the kinetic energy at the turning point is
still less than $mc^2$, so the loss function of Equation (3) can be used.

Consider interstellar CRs with the density in momentum space having the same power-law slope for both relativistic and
non-relativistic energies. The corresponding kinetic energy spectrum reads
\begin{equation}\label{j_rel}
j_i(\mathscr{E})=j_0\left(\frac{2mc^2\mathscr{E}_0}{\mathscr{E}^2+2mc^2\mathscr{E}}\right)^a.
\end{equation}
For non-relativistic particles, Equation \eqref{j_rel} is reduced to the spectrum of Equation (2), adopted in the paper. In the
ultra-relativistic regime, the spectrum becomes $j_i(\mathscr{E})=j_{0,\rm rel}(\mathscr{E}_0/\mathscr{E})^{2a}$ with
$j_{0,\rm rel}=(2mc^2/\mathscr{E}_0)^aj_0$. This allows us to easily extend our calculations to the ultra-relativistic case.

Following the same logic as in the main text, but using $\mathscr{E} = pc$ instead of $\mathscr{E} = p^2/(2m)$, we find
\begin{equation}
j(\mathscr{E}_\parallel, \mathscr{E}_\perp, z)=2\pi \mathscr{E_\perp} \frac{j_i(\mathscr{E} + e|\phi|)}{(\mathscr{E} + e|\phi|)^2}\;,
\end{equation}
with $\mathscr{E} =\sqrt{\mathscr{E}_\|^2 + \mathscr{E}_\perp^2}$, in lieu of Equation (11). While Equation (12) remains unchanged,
the calculation of the initial $\mathscr{E}_{\perp}^{\rm cr}$ proceeds differently. We assume that the critical energy {\it
near the turning point} is non-relativistic. Then its value is still given by Equation (15). Next, we note that the perpendicular
momentum $p_\perp$ is a conserved quantity (neglecting losses). Setting $p_\perp^2/(2m)$ at the turning point equal to the
RHS of Equation (15), we find the critical value of the {\it initial} transverse energy $\mathscr{E}_\perp=cp_\perp$,
\begin{equation}
\mathscr{E}_\perp^{\rm cr} = \sqrt{2mc^2\mathscr{E}_0} \left(M_{\rm cr}
\frac{nL_0}{eE_0}\right)^{\frac{1}{2s}} \left(\frac{\mathscr{E}_\|}{\mathscr{E}_0}\right)^\frac{1-f_{\rm rel}}{2sf_{\rm rel}},
\label{eq:eperpcrit}
\end{equation}
to be substituted in Equation (12). As before, we approximate $\mathscr{E}_\perp^{\rm cr}\ll \mathscr{E}_\|$ and obtain an
equation analogous to Equation (16), which yields
\begin{equation}
f_{\rm rel} = \frac{1+s}{1+2s+2as}\;,
\end{equation}
and
\begin{equation}
\frac{n z_{0,\rm rel}}{N_0} = \left[\frac{(1+s)^{\frac{1}{s}-1} f_{\rm rel}^{\frac{1}{s} +1}}{M_{\rm cr}^\frac{1}{s}}
\frac{2as\sigma n}{\pi e^2j_{0,\rm rel}N_0}\left(\frac{\mathscr{E}_0}{2mc^2}\right) \right]^{\frac{s}{1+s}}.
\label{eq:ngz0}
\end{equation}
Equation \eqref{eq:ngz0} is similar to Equation (18) where parameters $a$, $f$, and $j_0$ are replaced with the respective
ultra-relativistic values, and the extra factor $\mathscr{E}_0/(2mc^2)$ originates from the square-root factor in
Equation \eqref{eq:eperpcrit}.

Using Equation \eqref{eq:ngz0}, we obtain an ultra-relativistic relation $e|\phi|_{\rm rel}/\mathscr{E}_{\rm ext}$ versus $N$. By
comparing this with the non-relativistic relation, Equation (19), we derive Equation (24) where $N_{\rm rel}$ is the column such that
\begin{equation}\label{psi}
\frac{e|\phi(N_{\rm rel})|}{mc^2}= 2\left[\frac12\left(\frac{1+2s+2as}{1+s+as}\right)^{\frac1s+1}\right]^{\frac1{1+a}}
\equiv\psi(a,s).
\end{equation}
For $1\leq a\leq 2$, we have $\psi(a,s)\approx2.4$.

Finally, we verify the assumption made in the beginning, that the kinetic energy near the turning point can still be
considered non-relativistic. To identify the column density $N_{\rm max}$ where the assumption breaks down, we use Equation (13)
with $\mathscr{E}_\perp^{\rm cr}=mc^2$, which gives the electric field $E_{\rm max}$ at that turning point. Substituting
this to $E_{\rm max}=f_{\rm rel}(n/N_{\rm max})|\phi_{\rm max}|$, which follows from Equations (9) and (10), we obtain $e|\phi_{\rm
max}|=(M_{\rm cr}/f_{\rm rel})N_{\rm max}L(mc^2)$. Next, we introduce the column density $N_*\approx3\times10^{22}$~cm$^{-2}$
at which the electron extinction energy in Equation (4) is equal to $mc^2$. Combining the two equations, we derive
\begin{equation}\label{phi_max}
\frac{e|\phi_{\rm max}|}{mc^2}=\frac{M_{\rm cr}}{f_{\rm rel}(1+s)}\:\frac{N_{\rm max}}{N_*}\;.
\end{equation}
Finally, by virtue of Equations (9) and \eqref{psi} we write $e|\phi_{\rm max}|/mc^2=\psi(N_{\rm max}/N_{\rm rel})^{f_{\rm rel}}$.
Equating with Equation \eqref{phi_max} gives
\begin{equation}\label{N_max}
\frac{N_{\rm max}}{N_{\rm rel}}=\left(\frac{\psi f_{\rm rel}(1+s)}{M_{\rm cr}}\:
\frac{N_*}{N_{\rm rel}}\right)^{\frac1{1-f_{\rm rel}}}.
\end{equation}
Note that $N_{\rm max}$ is comparable to, or larger than $N_*$.
For the conditions illustrated in Figure 1, $N_{\rm rel}\approx 3 \times 10^{21}~{\rm cm^{-2}}$ ($2 \times 10^{21}~{\rm
cm^{-2}}$) for $a = 1$ (2), resulting in $N_{\rm max} \approx 5 \times 10^{22}~{\rm cm^{-2}}$ ($2 \times 10^{22}~{\rm
cm^{-2}}$). Thus, $N_{\rm max}\gg N_{\rm rel}$ and our assumption is well justified for molecular clouds.

\section{Appendix D: Joule Heating}
\label{sect:Joule}

We calculate the ratio of Joule heating $H_J$, given by Equation~\eqref{eq:HJ}, to regular gas heating $H_{\rm CR}$ by
CRs, given by Equation \eqref{eq:HI}. Substituting Equation~(\ref{eq:finalZeta}) into Equation~\eqref{eq:HI}, we obtain
\begin{equation}
H_{\rm CR} = 4\pi \eta B j_0 L_0 \mathscr{E}_0n \left(\frac{e |\phi|}{\mathscr{E}_0}\right)^{-(a+s-1)}.
\end{equation}
Inserting $E=f|\phi|/z$ in Equation~\eqref{eq:HJ} and keeping in mind that $\mathscr{E}_0/L_0=(1+s)N_0$, we can then write
the ratio as
\begin{equation}
\frac{H_J}{H_{\rm CR}} = \frac{(1+s)f^2}{4\pi \eta B}\:\frac{\sigma n}{e^2j_0N_0}\left(\frac{N_0}{nz_0}\:\frac{z_0}{z}\right)^2
\left(\frac{e|\phi|}{\mathscr{E}_0}\right)^{a+s+1}.
\end{equation}
Next, by virtue of Equations~\eqref{eq:Eext} and \eqref{eq:phi0} this can be written as
\begin{equation}
\frac{H_J}{H_{\rm CR}} = Q \left(\frac{N_0}{nz_0}\right)^{\frac{s-1}{s}}\left(\frac{N}{N_0}\right)^\frac{1+a+s}{1+s}
\left(\frac{z_0}{z}\right)^2\left( \frac{e |\phi|}{\mathscr{E}_{\rm ext}}\right)^{a+s+1},
\end{equation}
where
\begin{equation}
Q = \frac{(1+s)^{2-\frac1s} f^{1-\frac1s} M_{\rm cr}^{\frac1s}}{4\eta as B}\;,
\end{equation}
is a function of $a$ and $s$ (for given $\eta$). Then, inserting $z_0/z = (e\phi/\mathscr{E}_0)^{-1/f}$ with $f$ from
Equation~\eqref{eq:csol}, we find
\begin{equation}
\frac{H_J}{H_{\rm CR}} = Q \left[\left(\frac{N_0}{nz_0}\right)\left(\frac{N}{N_0}\right)^\frac{s(1+s-a)}{(1+s)^2}
\left(\frac{e |\phi|}{\mathscr{E}_{\rm ext}}\right)^\frac{s(1+s-a)}{1+s}\right]^\frac{s-1}{s}.
\end{equation}
We notice that, using Equation~\eqref{phi_to_E}, the first two factors in the brackets can be expressed via $e |\phi|/
\mathscr{E}_{\rm ext}$. This finally yields
\begin{equation}
\frac{H_J}{H_{\rm CR}} = Q\left(\frac{e|\phi|}{\mathscr{E}_{\rm ext}} \right)^{-\frac1s+s}.
\end{equation}
For parameters of Figure~\ref{ionizationRates}, $Q$ varies between about 7 and 10 for $1\leq a \leq 2$.


\end{document}